\documentclass[12pt]{iopart}
 
\usepackage{iopams}
\expandafter\let\csname equation*\endcsname\relax
\expandafter\let\csname endequation*\endcsname\relax
\usepackage{amsmath} 
\usepackage{graphicx} 
\usepackage{dcolumn}% Align table columns on decimal point
\usepackage{bm}% bold math
\usepackage{subcaption}
\usepackage{float}
\usepackage{lineno}
\usepackage{xcolor}  
\expandafter\let\csname equation*\endcsname\relax
\expandafter\let\csname endequation*\endcsname\relax
\RequirePackage[colorlinks=true, citecolor=red, urlcolor=blue, linkcolor=blue]{hyperref}

\begin{document}
%============================ Title ====================
\title[]{A feasibility study of the light front analysis of ultrarelativistic nuclear collisions in ALICE at LHC}
\author{Rahul Ramachandran Nair} 
\address{National Centre For Nuclear Research, 02-093 Warsaw, Poland.\footnote{This work was performed while the author was affiliated with National Centre For Nuclear Research, Warsaw, Poland. Present address: Sreeraj Bhavan, Karukachal PO, Kottayam, Kerala, India-686540}}
\ead{\mailto{physicsmailofrahulnair@gmail.com}}
\vspace{10pt}
\begin{indented}
\item[]\today
\end{indented} 
%============================ Abstract  ====================

\begin{abstract}
An analysis involving the light front variables is performed over the inclusively produced $\pi^{\pm}$, $K^{\pm}$ and $p(\bar{p})$ in the Pb-Pb collisions simulated using the UrQMD event generator at $\sqrt{s} = 2.76$ TeV to study the possible formation of a thermalised medium in these collisions. It is demonstrated that there exist surfaces defined by constant values of the light front variable in the phase space of these hadrons which can select a group of thermalised particles. The temperatures are extracted for each species and are compared with those obtained from a similar calculation involving the kinematical and acceptance restrictions which are typically bound to the collider experiments. It is shown that the analysis is feasible to perform with the data collected at the ALICE experiment at LHC. The analysis may be considered as a prototypical study to a similar experimental analysis of the data from the heavy-ion collisions at RHIC and LHC after implementing the necessary corrections and calibrations relevant to the experimental apparatuses.
\end{abstract}

%============================ Introduction =============================================
\section{Introduction} 
The scale-invariant light front variable was proposed in \cite{Garsevanishvili78, Garsevanishvili79} for studying the particle production in hadron-hadron and nucleus-nucleus interactions. The analysis scheme based on these variables were implemented for a variety of hadron-hadron,  nucleus-nucleus interactions at various centre of mass energies for charged pions in \cite{Amaglobeli99, Djobava03, Chkhaidze2006} etc. The main conclusion of these studies was that the phase space of the secondary pions is divided into two parts with significantly different characteristics with one among the groups following the Boltzmann statistics corresponding to thermal equilibrium. The proposed scale and Lorentz invariant light front variable takes the following form in the centre of the mass frame
\begin{equation} \label{EqnXi}
\xi^{\pm} = \pm \frac{E \pm p_{z}}{\sqrt{s}} = \pm  \frac{E + |p_{z}|}{\sqrt{s}}
\end{equation}
where, $s$ is the Mandelstam variable, $p_{z}$ is the z-component of the momentum and E is the energy of the particle. The positive and negative signs in equation \eqref{EqnXi} corresponds to the two hemispheres. It was observed in the analysis performed in  \cite{Amaglobeli99, Djobava03, Chkhaidze2006} that there is a maximum in the $|\xi^{\pm}|$ distribution of charged pions, produced in the hadron-hadron and nucleus-nucleus interactions, around a small value of $|\xi|$ denoted by $|\tilde{\xi}|$. The charged pions contributing to the left and right side of this maximum were observed to have distinct features. Those pions with $|\xi| < |\tilde{\xi}|$ had a flat angular distribution with respect to a sharply anisotropic angular distribution of those particles with $|\xi| > \tilde{\xi}$. The slopes of the square of the transverse momentum distribution of these two sets of particles also were found to be different. Hence it was concluded that the $\pi^{\pm}$ mesons with $|\xi| < |\tilde{\xi}|$ might have reached thermal equilibrium. To enlarge the scale in the region of smaller $|\xi|$ values, a convenient variable $\zeta$ can be defined as follows:
\begin{equation}\label{zetadef}
\zeta^{\pm} =     \mp \ln(|\xi^{\pm}|)
\end{equation}
A simple statistical model was used in \cite{Amaglobeli99, Djobava03} to test the hypothesis of thermalisation in the low energy collisions with these light front variables. The kind of analysis is termed as the 'light front analysis'. To perform such a study with the actual experimental data collected from the heavy-ion collisions at the LHC, one needs to consider various kinematical and acceptance cuts bound to the experiment. The successful performance of the light front scheme in describing the LHC data would confirm that at least a group of hadrons in the heavy-ion collisions at LHC got indeed thermalised. In the following section, we will revisit the details and obtain the equations needed for our analysis. The scheme of light front analysis, the treatment of kinematical and acceptance constraints coming from experiments are discussed in the later sections. We will demonstrate using the lead-lead collisions simulated with the UrQMD event generator at $\sqrt{s} = 2.76$ TeV that the light front variable based scheme of analysis can be effectively implemented for testing the existence of a thermalised medium in the heavy-ion collisions at LHC.
%========================================================================================
\section{Analytic expressions for the spectra} 
The analytic expressions for the description of the spectrum of light front variable $\zeta^{\pm}$, the polar angle $cos(\theta)$ and the square of the transverse momentum $p_T^2$ which are used for the light front analysis in \cite{Amaglobeli99, Djobava03} are revisited and derived in this section. The invariant differential cross-section in terms of the light front variables can be written as follows:
\begin{equation}\label{cross}
E\frac{d \sigma}{d \textbf{p}} = \frac{\xi^{\pm}}{\pi}\frac{d^2\sigma}{d\xi^{\pm}dp_T^2} = \frac{1}{\pi}\frac{d^2\sigma}{d\zeta^{\pm}dp_T^2}
\end{equation}
A scheme of analysis can be constructed to study the thermalisation if we assume that the energy distribution of the particles is of the form 
\begin{equation}\label{eqboltz}
f(E) \sim  exp(-E/T)
\end{equation} 
which corresponds to a system that has reached thermal equilibrium \cite{Kardar}. Let us suppose that the particles under consideration has reached thermal equilibrium and hence one can write
\begin{equation}\label{modelcross1}
\frac{d \sigma}{d \textbf{p}} = f(E)
\end{equation}
With this assumption, one can obtain the expression for the $\zeta^{\pm}$ distribution using the form of the invariant differential cross section in equation \eqref{cross} as follows:
\begin{equation}
\frac{dN}{d\zeta^{\pm}} \sim \int_0^{p_T^2(max)} E f(E)dp_T^2
\label{ZetaInt}
\end{equation}
where $p_{T,max}^2$ is given by
\begin{equation}
p_{T,max}^2 = (\xi^{\pm}\sqrt{s})^{2} - m^{2} \label{ptmax}
\end{equation}
which is the maximum value of $p_T^2$ of the particle with mass $m$, kinematically allowed for a specific value of $\xi$ ( or $\zeta$) in a collision with the available centre of mass energy being $\sqrt{s}$. The value of $\zeta^{\pm}$ down to which equation \eqref{ZetaInt} can be used to describe the experimental $\zeta^{\pm}$ distribution is taken as $\tilde{\zeta^{\pm}}$. Note that a constant value of $\zeta^{\pm}$ defines a paraboloid in the phase space of the inclusively produced particles in heavy-ion collisions. As per the proposed construction, the particles falling under the paraboloid defined by $\tilde{\zeta^{\pm}}$ supposedly forms a thermalised group. Now let us consider the invariant cross section in equation \eqref{cross} again. It can be written in terms of the polar angle $\theta$, azimuthal angle $\phi$, and total momentum $p$ of the particle as follows:
\begin{equation}\label{modelcross2}
E\frac{d \sigma}{d \textbf{p}} = E\frac{1}{p^2} \frac{d^3 \sigma}{d\phi dp dcos(\theta)}
\end{equation}
If we assume that the particles are thermalised, then it follows that
\begin{equation}\label{modelcross2b}
\frac{d\sigma}{dcos(\theta)} = \int_0^{p_{max}}\int_0^{2\pi} p^2 f(E) d\phi dp
\end{equation}
With an integration over $\phi$, one can obtain the following expression for the cos($\theta$) distribution of particles:
\begin{equation}
\frac{dN}{dcos(\theta)} \sim \int_0^{p_{max}} f(E) p^2dp
\label{CosInt}
\end{equation}
The value of $p_{max}$ in equation \eqref{CosInt} is determined by the specific region in the phase space we are interested in. From the consideration of the $\zeta^{\pm}$ distribution, we have a closed boundary in the phase space inside which the particles are supposedly thermalised. This boundary is uniquely defined by the value of $\tilde{\zeta^{\pm}}$. If we consider the very same region in the phase space, then the $p_{max}$ in equation \eqref{CosInt} will have the following form:
\begin{equation}
p_{max}  = \frac{-\tilde{\xi^{\pm}}\sqrt{s}cos(\theta) + \sqrt{(\tilde{\xi^{\pm}}\sqrt{s})^2  - m^2 sin^2(\theta)}}{sin^2(\theta)}
\label{pmax}
\end{equation}
If the particles inside the paraboloid defined by $\tilde{\zeta^{\pm}}$ in the phase space are thermalised, then it follows that the polar angle distribution of those particles can be described using equation \eqref{CosInt} with $p_{max}$ given by equation \eqref{pmax}. We can as well write the cross-section in terms of rapidity $y$ and $p_{T}^2$ of the particle as follows:
\begin{equation}\label{modelcross3}
E\frac{d \sigma}{d \textbf{p}} = \frac{d^2 \sigma}{dy dp_{T}^2}
\end{equation}
For a thermalised system, equation \eqref{modelcross3} can be written as 
\begin{equation}\label{modelcross4}
f(E) = \frac{d^2 \sigma}{dp_{z} dp_{T}^2}
\end{equation}
using the relation $dy = dp_{z}/E$. Hence the $p_{T}^2$ distribution of the particles will have the following form:
\begin{equation}
\frac{dN}{dp_T^2} \sim \int_0^{p_{z,max}} f(E)dp_{z}
\label{PtSqInt}
\end{equation}
If we make the similar step as we made for the angular distribution and consider the  $p_{T}^2$ distribution of the particles inside the boundary defined by $\tilde{\zeta^{\pm}}$, then the $p_{z,max}$ in equation \eqref{PtSqInt} is given by the expression
\begin{equation}
p_{z,max}  = \frac{m^2 + p_T^2 - (\tilde{\xi^{\pm}}\sqrt{s})^2}{-2\tilde{\xi^{\pm}}\sqrt{s}}
\label{pzmax}
\end{equation}
The particles inside the paraboloidal surface of $\zeta^{\pm} = \tilde{\zeta^{\pm}}$ must have a $\zeta^{\pm}$ distribution describable with equation \eqref{ZetaInt}, polar angle distribution describable with equation \eqref{CosInt} and  $p_{T}^{2}$ distribution describable with equation \eqref{PtSqInt} while the respective limits of integration are given by equation \eqref{ptmax}, equation \eqref{pmax} and equation \eqref{pzmax} for this construction to make sense. It can be used to test the presence of thermalisation in the heavy-ion collisions at LHC. For an analysis of the actual experimental data, one needs to take into account various restrictions imposed by the detectors. In the next section, a scheme for the implementation of the light front analysis of hadrons produced in the ultrarelativistic heavy-ion collisions and a method of incorporating the experimental constraints is described and performed with the events simulated using the UrQMD event generator.
%========================================================================================
\section{Scheme of analysis and the results} 
%------------------------- Zeta Plots ----------------------------------------------------
\begin{figure*}
	
	\begin{minipage}{.5\linewidth}
	
		\subfloat[$|\zeta^{\pm}|$($\pi^{\pm}$; ${\zeta}_C= 9.30$; without cuts)]{\includegraphics[scale=.43]{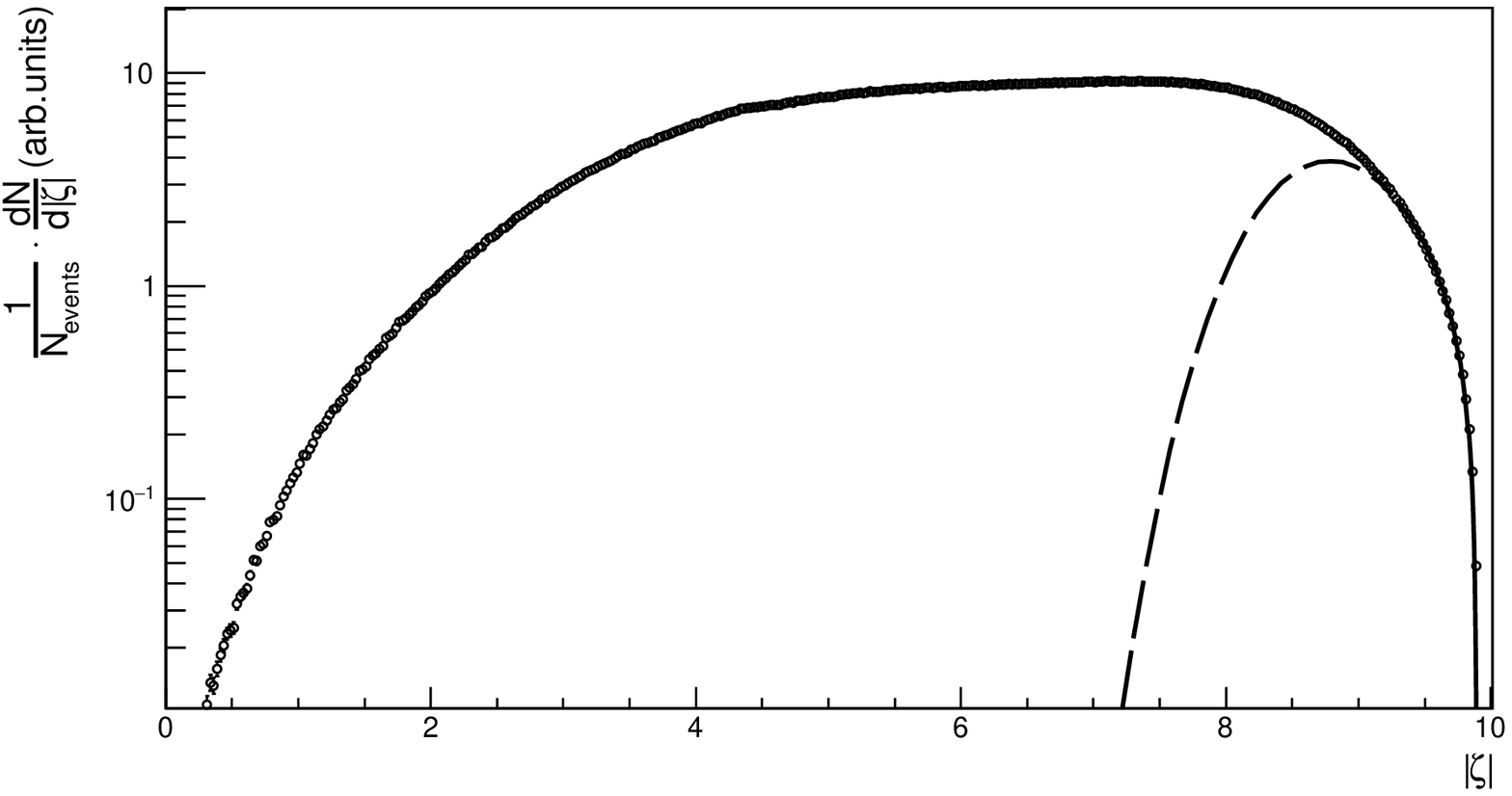}}
	\end{minipage}%
	\begin{minipage}{.5\linewidth}

		\subfloat[$|\zeta^{\pm}|$($\pi^{\pm}$; ${\zeta}_C= 9.30$; with cuts)]{\includegraphics[scale=.43]{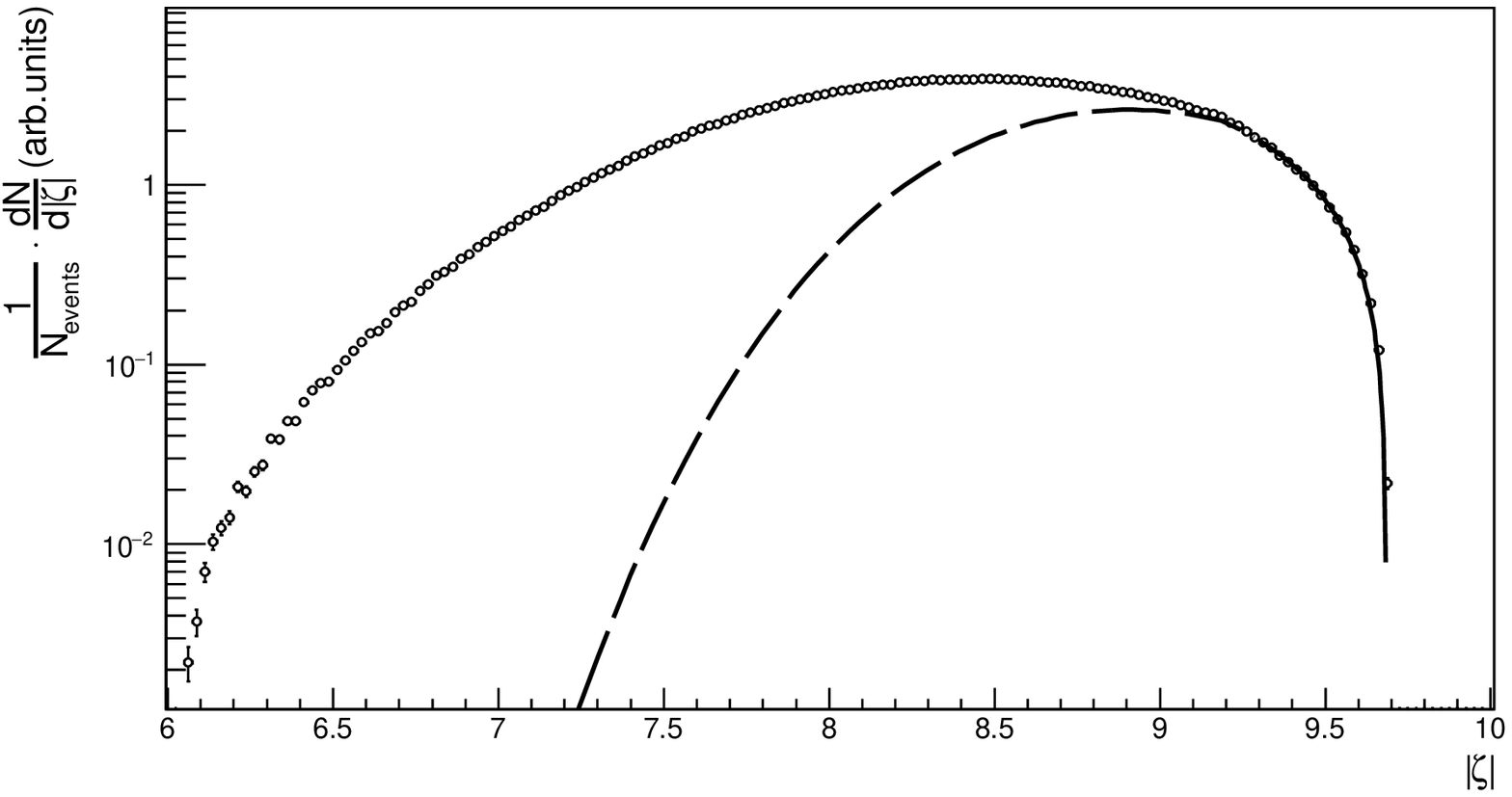}}
	\end{minipage}\\
		\begin{minipage}{.5\linewidth}
	
		\subfloat[$|\zeta^{\pm}|$($K^{\pm}$; ${\zeta}_C= 8.25$; without cuts)]{\includegraphics[scale=.43]{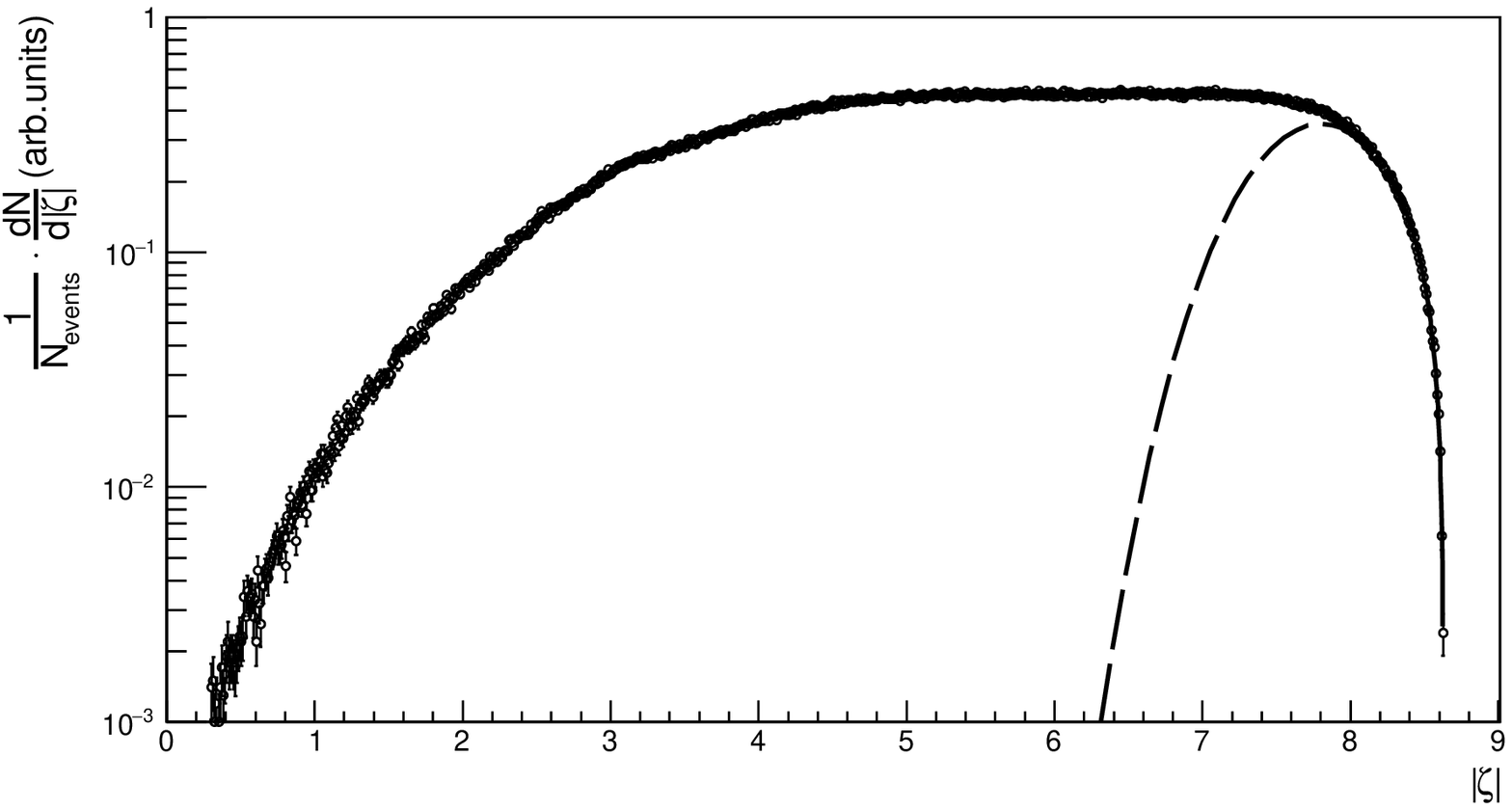}}
	\end{minipage}%
	\begin{minipage}{.5\linewidth}
	
		\subfloat[$|\zeta^{\pm}|$($K^{\pm}$; ${\zeta}_C= 8.25$; with cuts)]{\includegraphics[scale=.43]{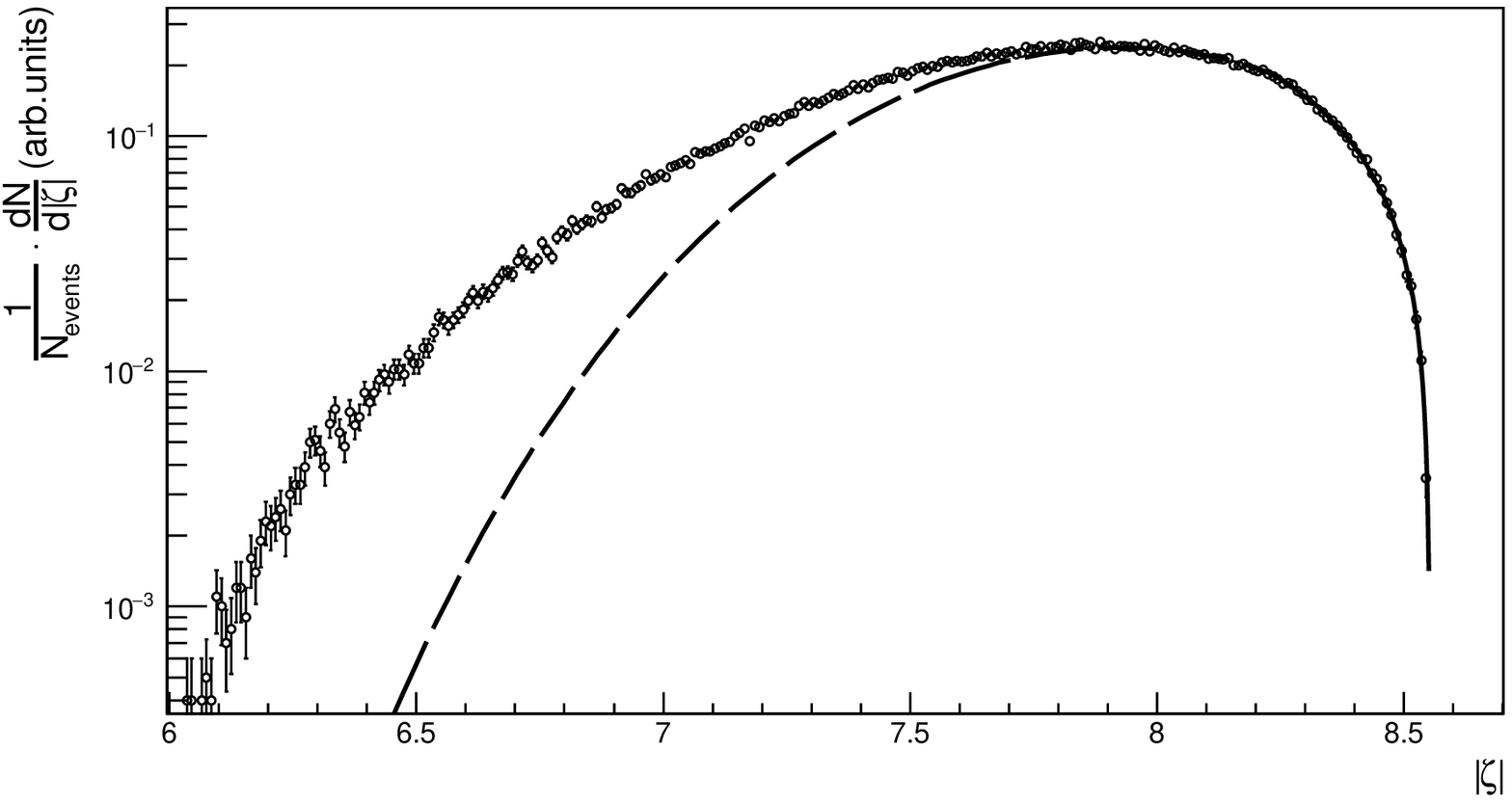}}
	\end{minipage}\\
		\begin{minipage}{.5\linewidth}
	
		\subfloat[$|\zeta^{\pm}|$($p(\bar{p})$; ${\zeta}_C= 7.60$; without cuts)]{\includegraphics[scale=.43]{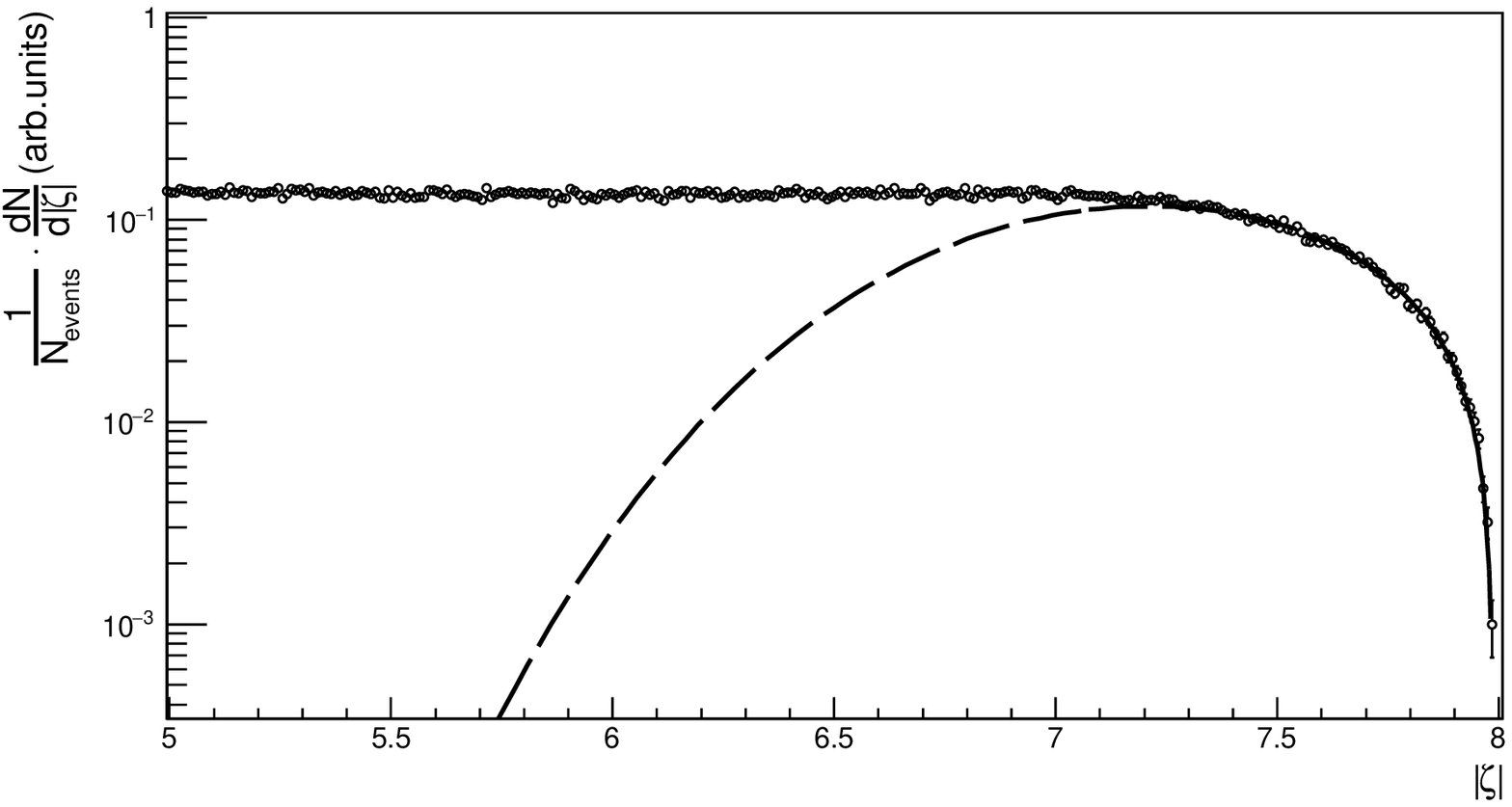}}
	\end{minipage}%
	\begin{minipage}{.5\linewidth}
	
		\subfloat[$|\zeta^{\pm}|$($p(\bar{p})$; ${\zeta}_C= 7.60$; with cuts)]{\includegraphics[scale=.43]{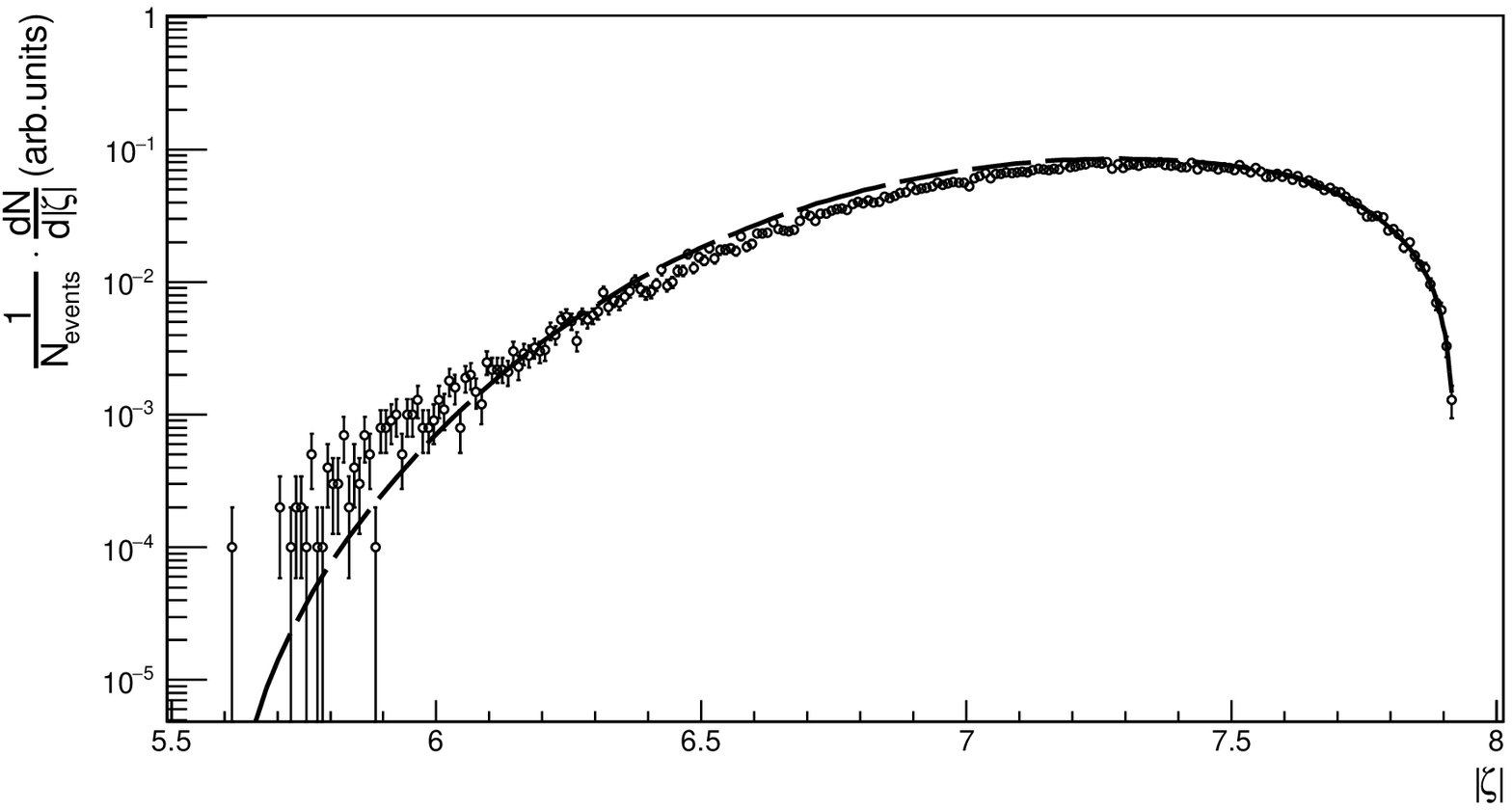}}
	\end{minipage}	
	\caption{$|\zeta^{\pm}|$ distribution of particles fitted with equation \eqref{ZetaInt} with and without kinematic and acceptance cuts. A full range $\zeta$ distribution of $p(\bar{p})$ can be seen in \ref{appendixA}.}
	\label{UrQMDZetaPlots}
\end{figure*}
%------------------------- Cos Plots ------------------------------------
\begin{figure*}
	
	\begin{minipage}{.5\linewidth}
	
		\subfloat[$|cos(\theta)|$ ($\pi^{\pm}$; ${\zeta}_C= 9.30$; without cuts)]{\includegraphics[scale=.43]{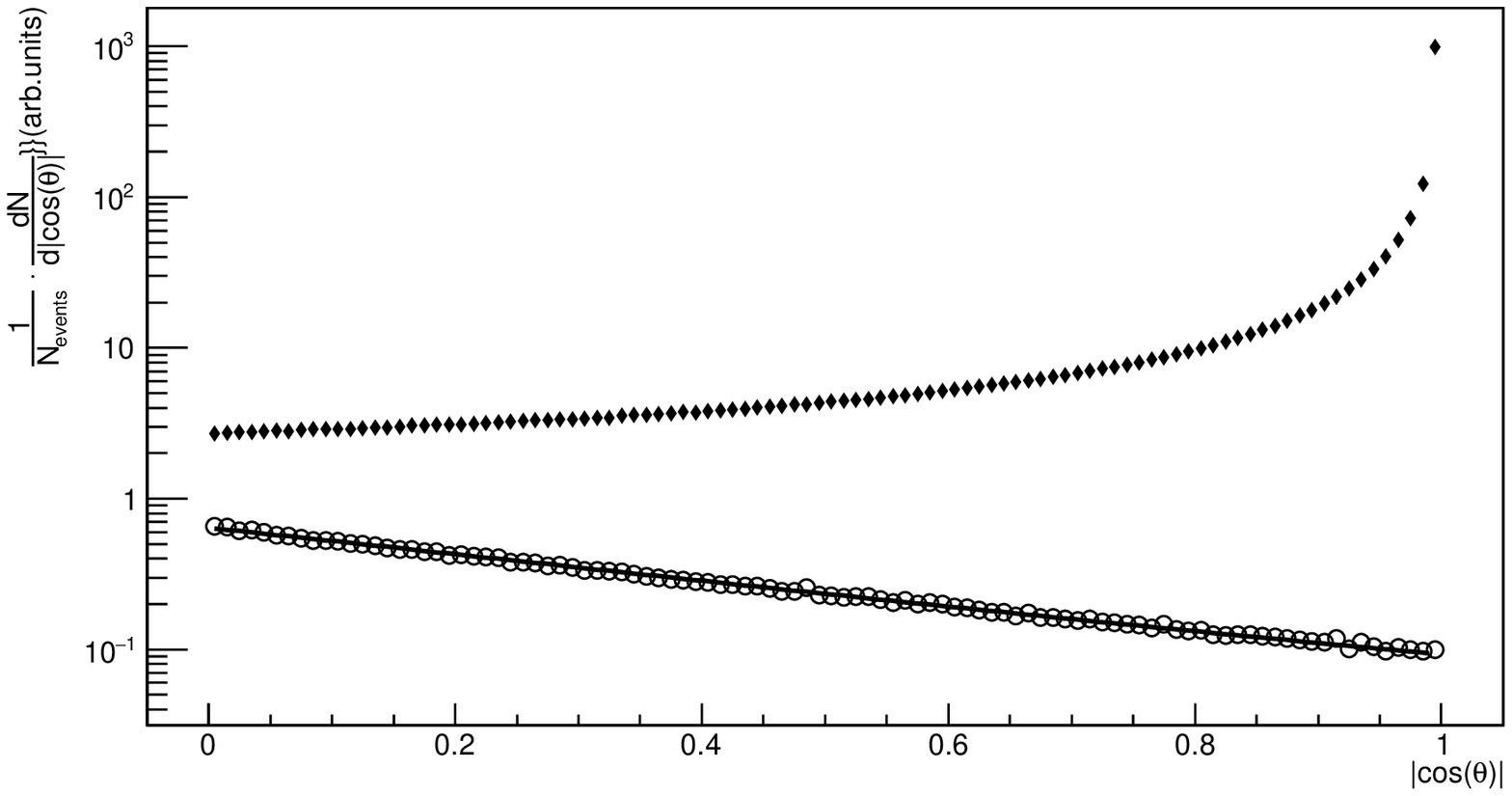}}
	\end{minipage}%
	\begin{minipage}{.5\linewidth}

		\subfloat[$|cos(\theta)|$ ($\pi^{\pm}$; ${\zeta}_C= 9.30$; with cuts)]{\includegraphics[scale=.43]{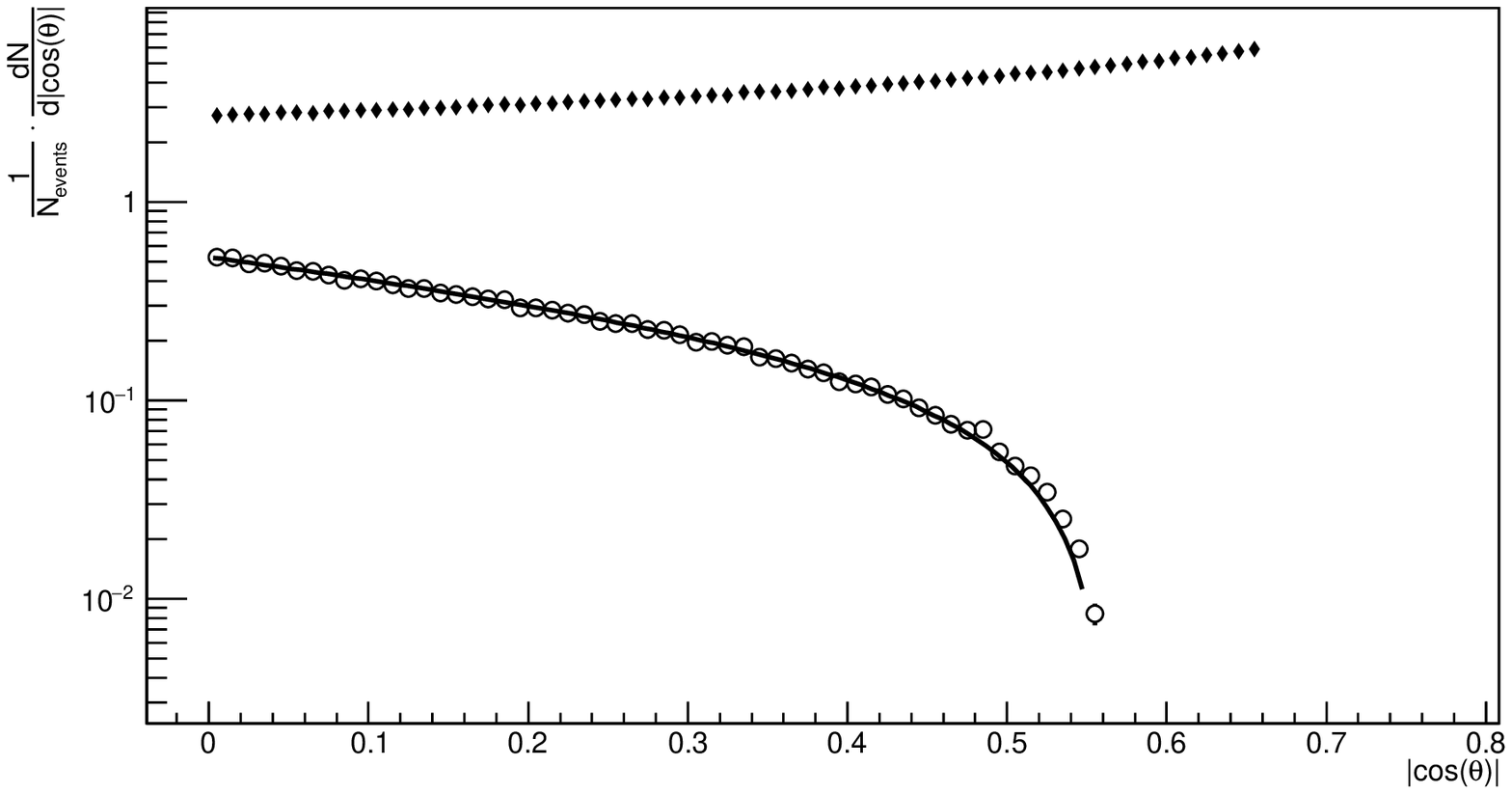}}
	\end{minipage}\\
		\begin{minipage}{.5\linewidth}
	
		\subfloat[$|cos(\theta)|$ ($K^{\pm}$; ${\zeta}_C= 8.25$; without cuts)]{\includegraphics[scale=.43]{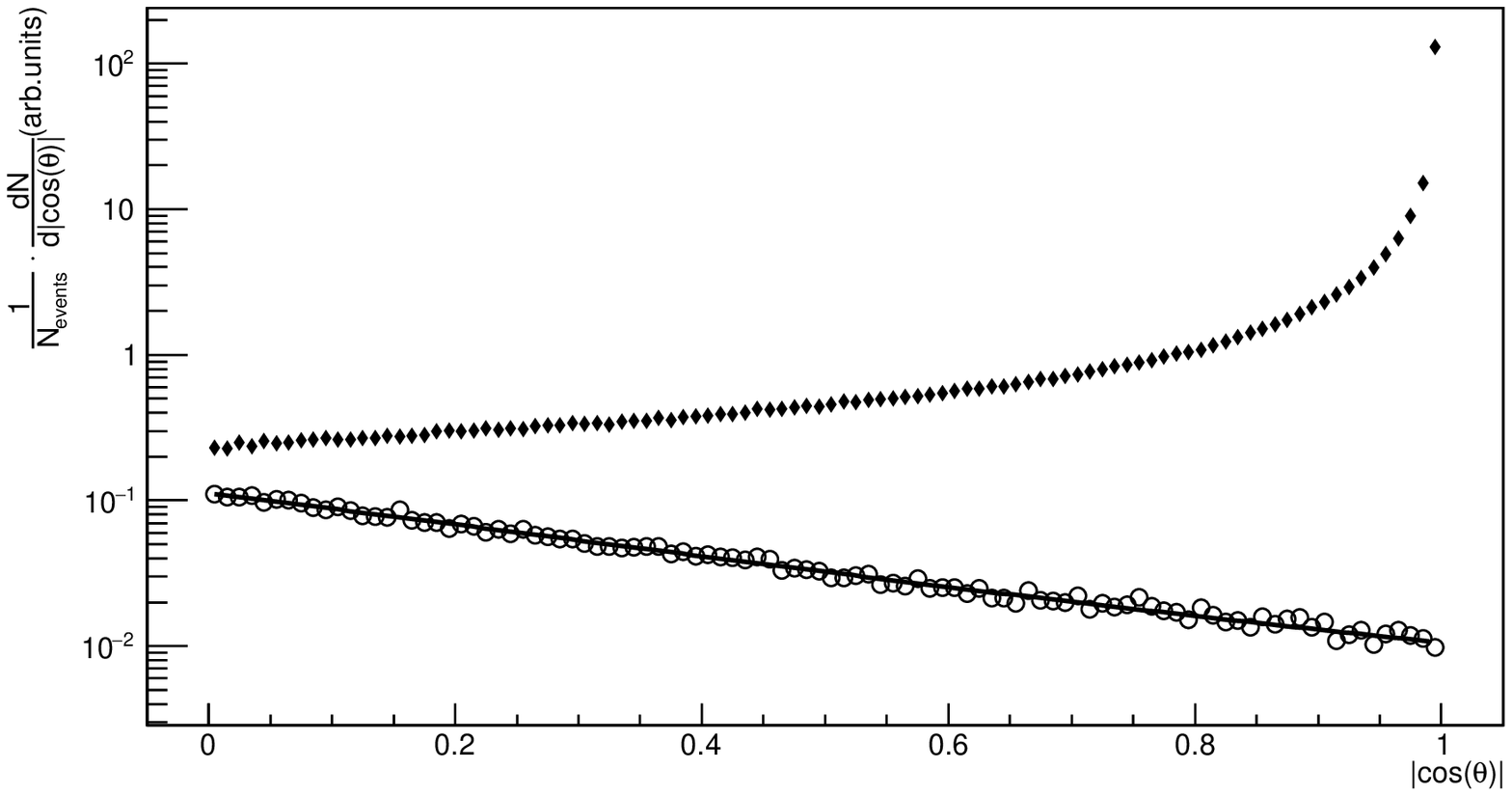}}
	\end{minipage}%
	\begin{minipage}{.5\linewidth}
	
		\subfloat[$|cos(\theta)|$ ($K^{\pm}$; ${\zeta}_C= 8.25$; with cuts)]{\includegraphics[scale=.43]{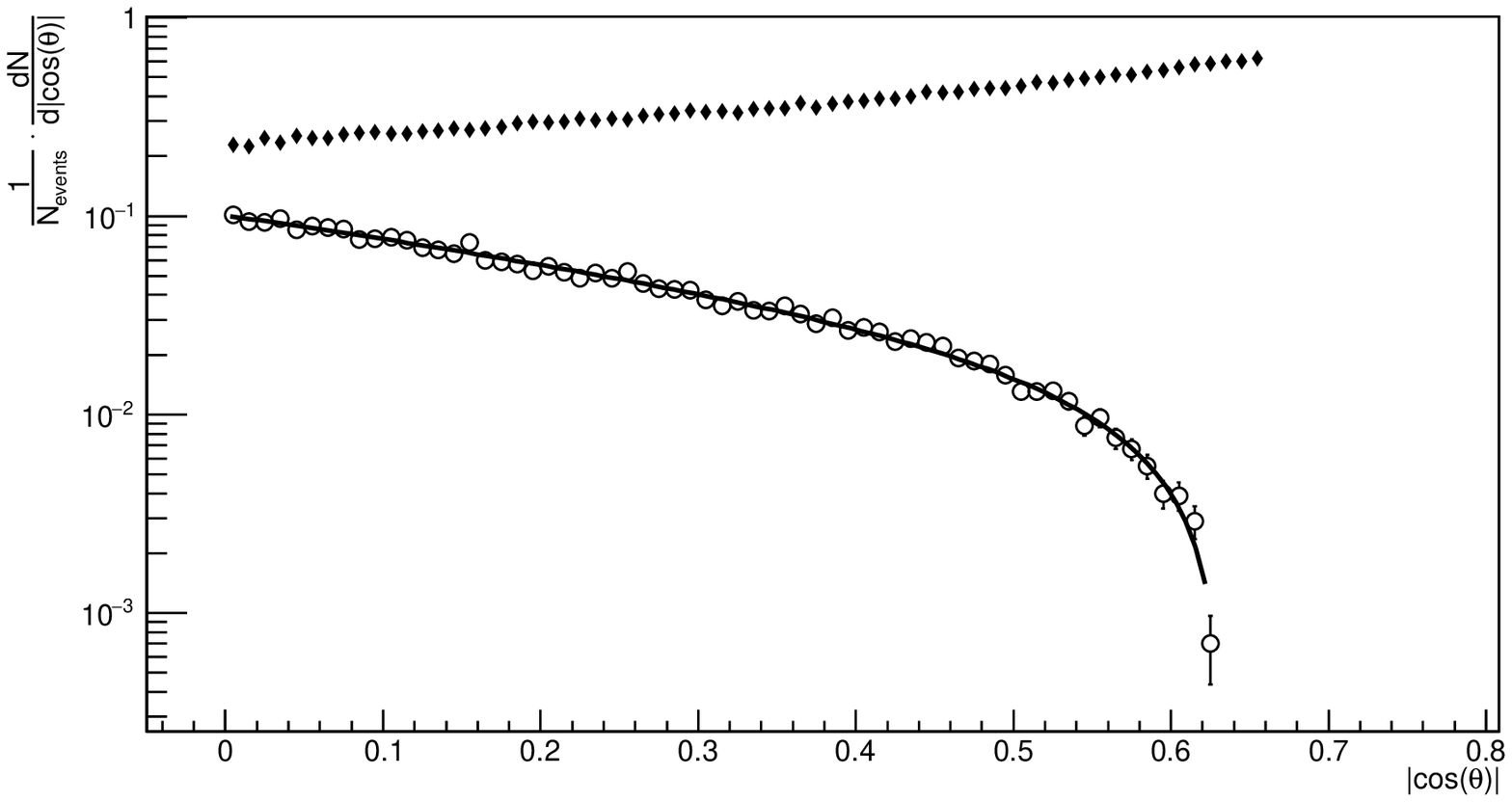}}
	\end{minipage}\\
		\begin{minipage}{.5\linewidth}
	
		\subfloat[$|cos(\theta)|$ ($p(\bar{p})$; ${\zeta}_C= 7.60$; without cuts)]{\includegraphics[scale=.43]{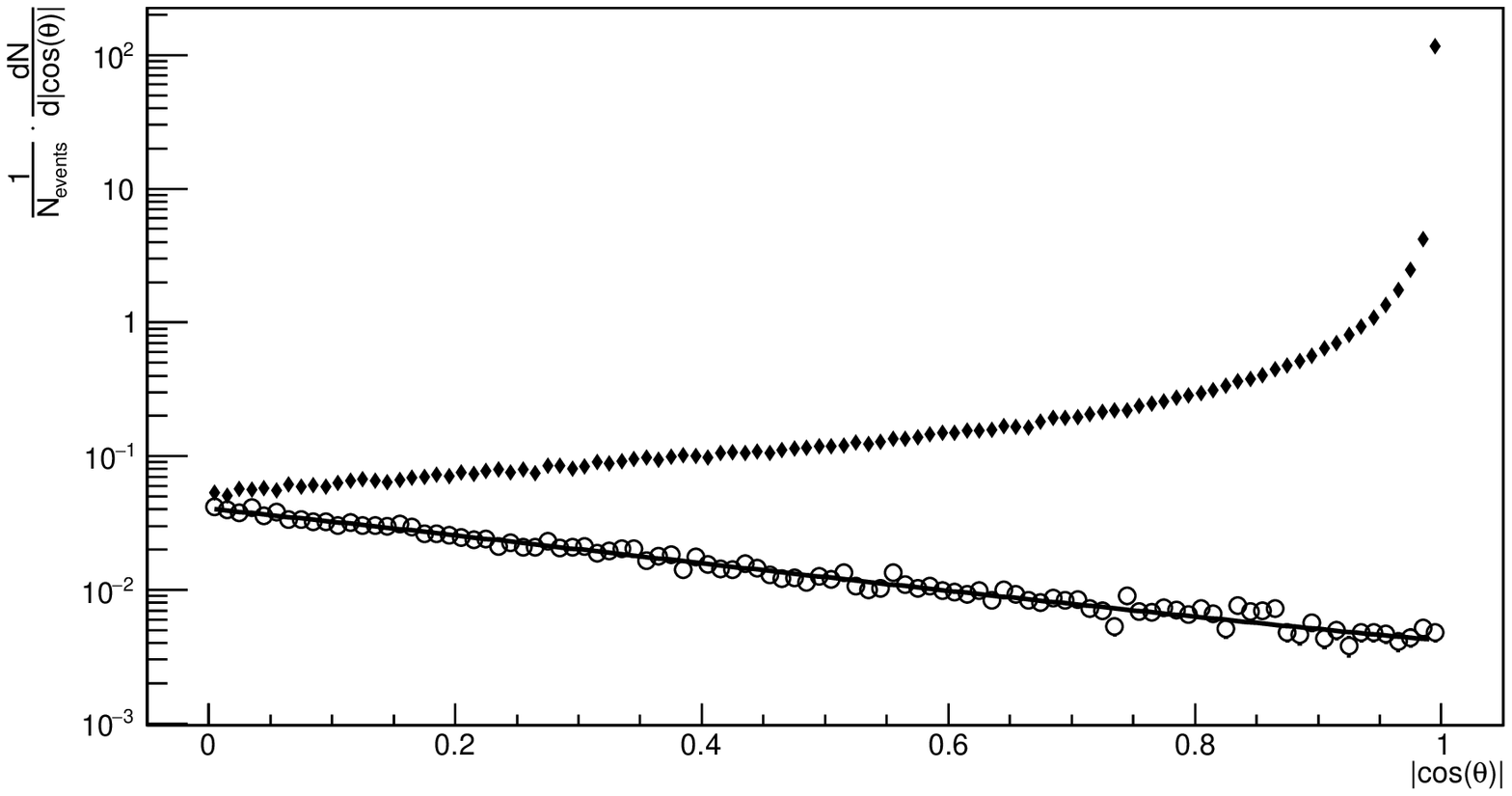}}
	\end{minipage}%
	\begin{minipage}{.5\linewidth}
	
		\subfloat[$|cos(\theta)|$ ($p(\bar{p})$; ${\zeta}_C= 7.60$; with cuts)]{\includegraphics[scale=.43]{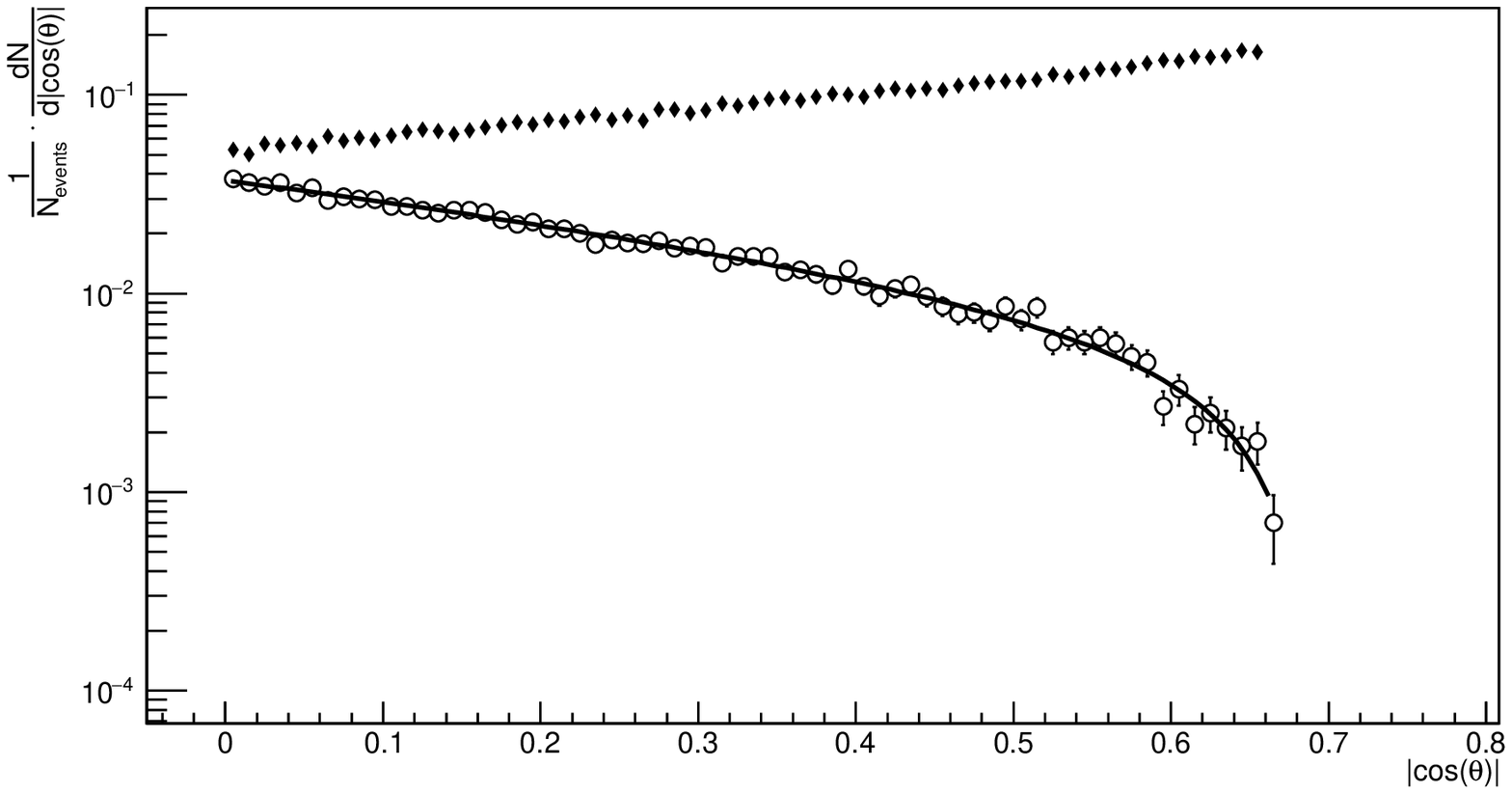}}
	\end{minipage}	
	\caption{The $|cos(\theta)|$ distribution of particles in Region -1 fitted with equation \eqref{CosInt} with and without kinematic and acceptance cuts anre shown in shown in open circles. Shown in filled diamonds are the corresponding distributions in Region-2.}
	\label{UrQMDCosPlots}
\end{figure*}
%------------------------- PtSq Plots -----------------------------------
\begin{figure*}
	
	\begin{minipage}{.5\linewidth}
	
		\subfloat[$p_{T}^{2}$ ($\pi^{\pm}$; ${\zeta}_C= 9.30$; without cuts)]{\includegraphics[scale=.43]{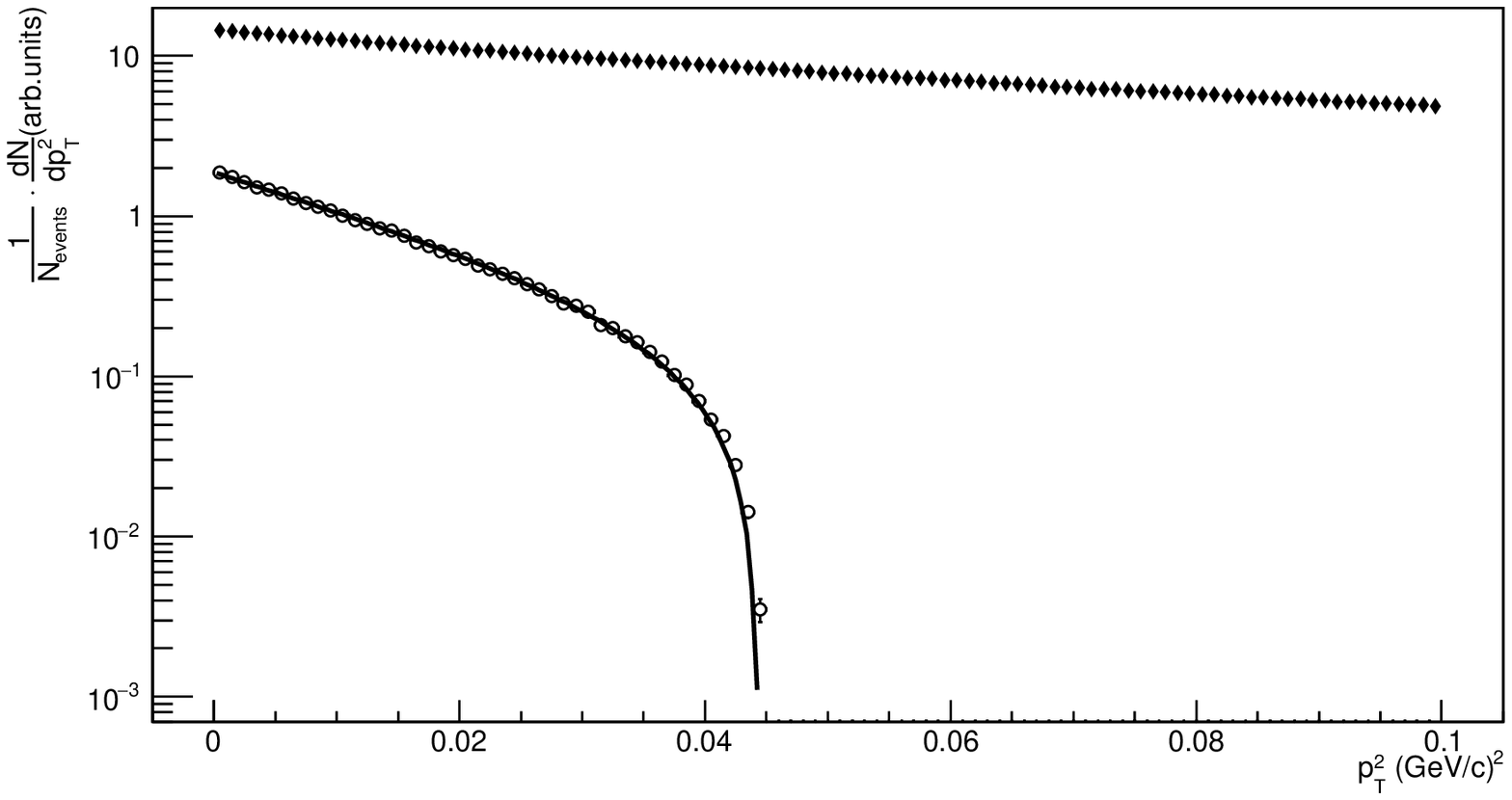}}
	\end{minipage}%
	\begin{minipage}{.5\linewidth}

		\subfloat[$p_{T}^{2}$ ($\pi^{\pm}$; ${\zeta}_C= 9.30$; with cuts)]{\includegraphics[scale=.43]{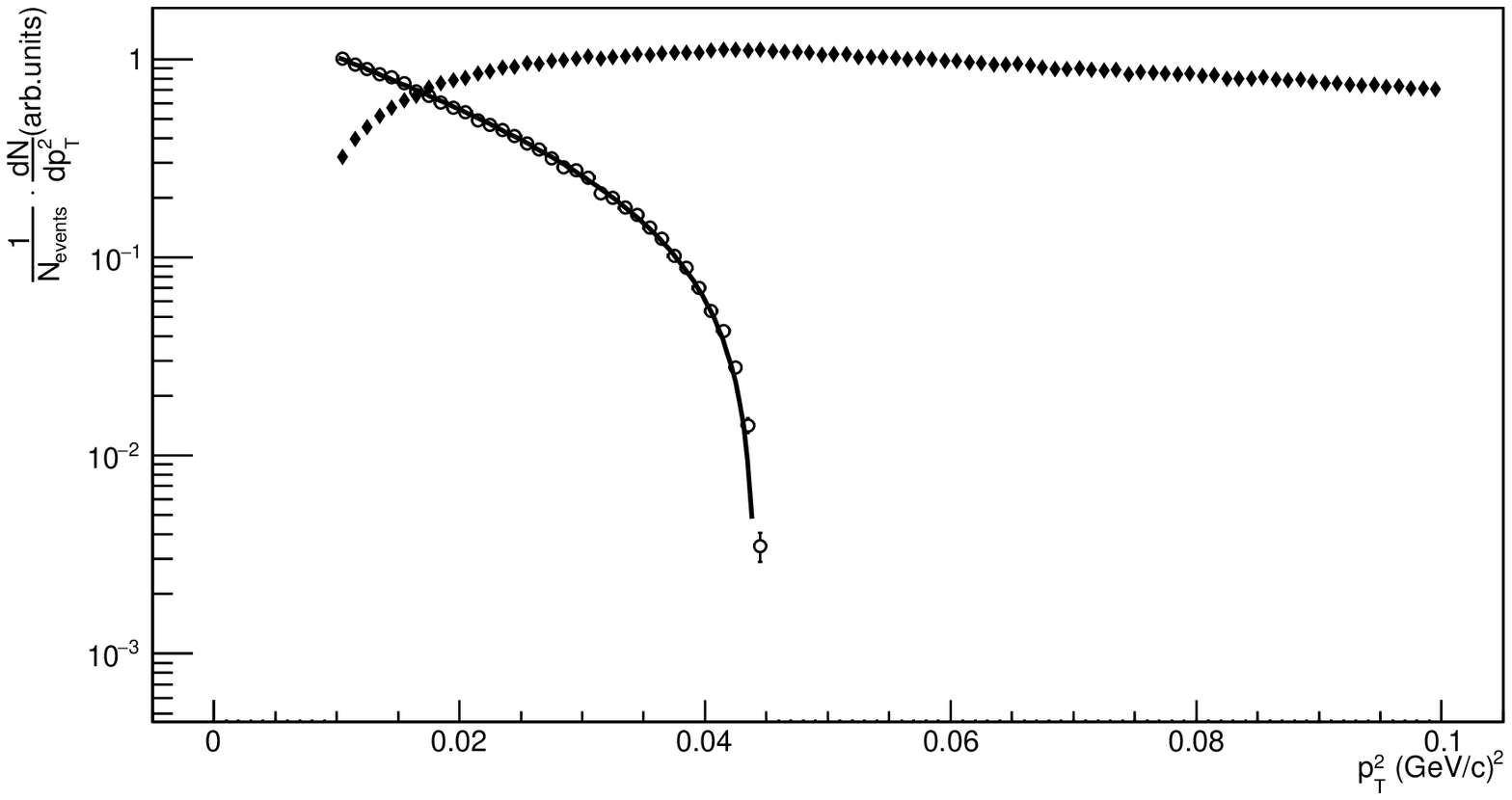}}
	\end{minipage}\\
		\begin{minipage}{.5\linewidth}
	
		\subfloat[$p_{T}^{2}$ ($K^{\pm}$; ${\zeta}_C= 8.25$; without cuts)]{\includegraphics[scale=.43]{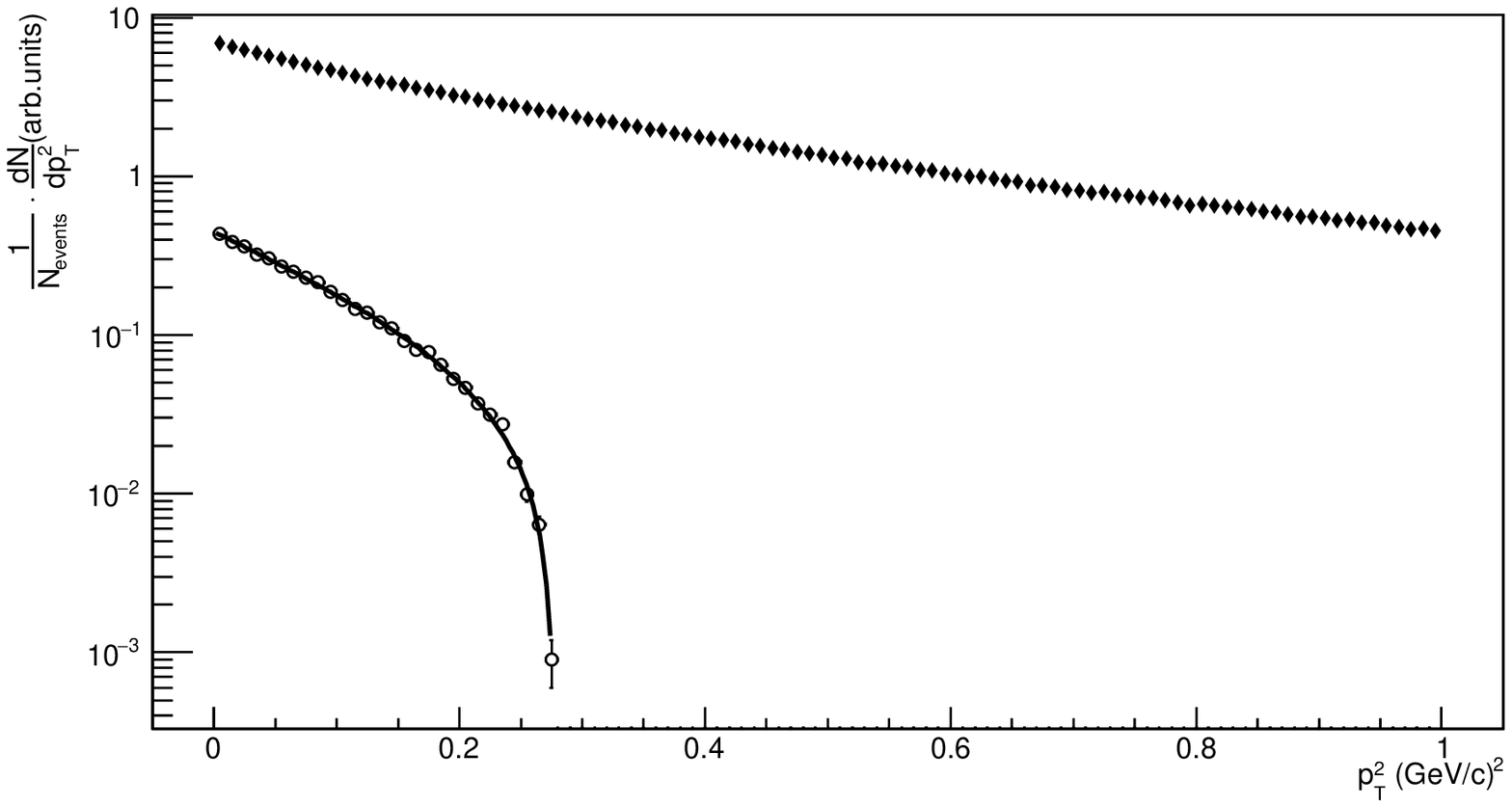}}
	\end{minipage}%
	\begin{minipage}{.5\linewidth}
	
		\subfloat[$p_{T}^{2}$ ($K^{\pm}$; ${\zeta}_C= 8.25$; with cuts)]{\includegraphics[scale=.43]{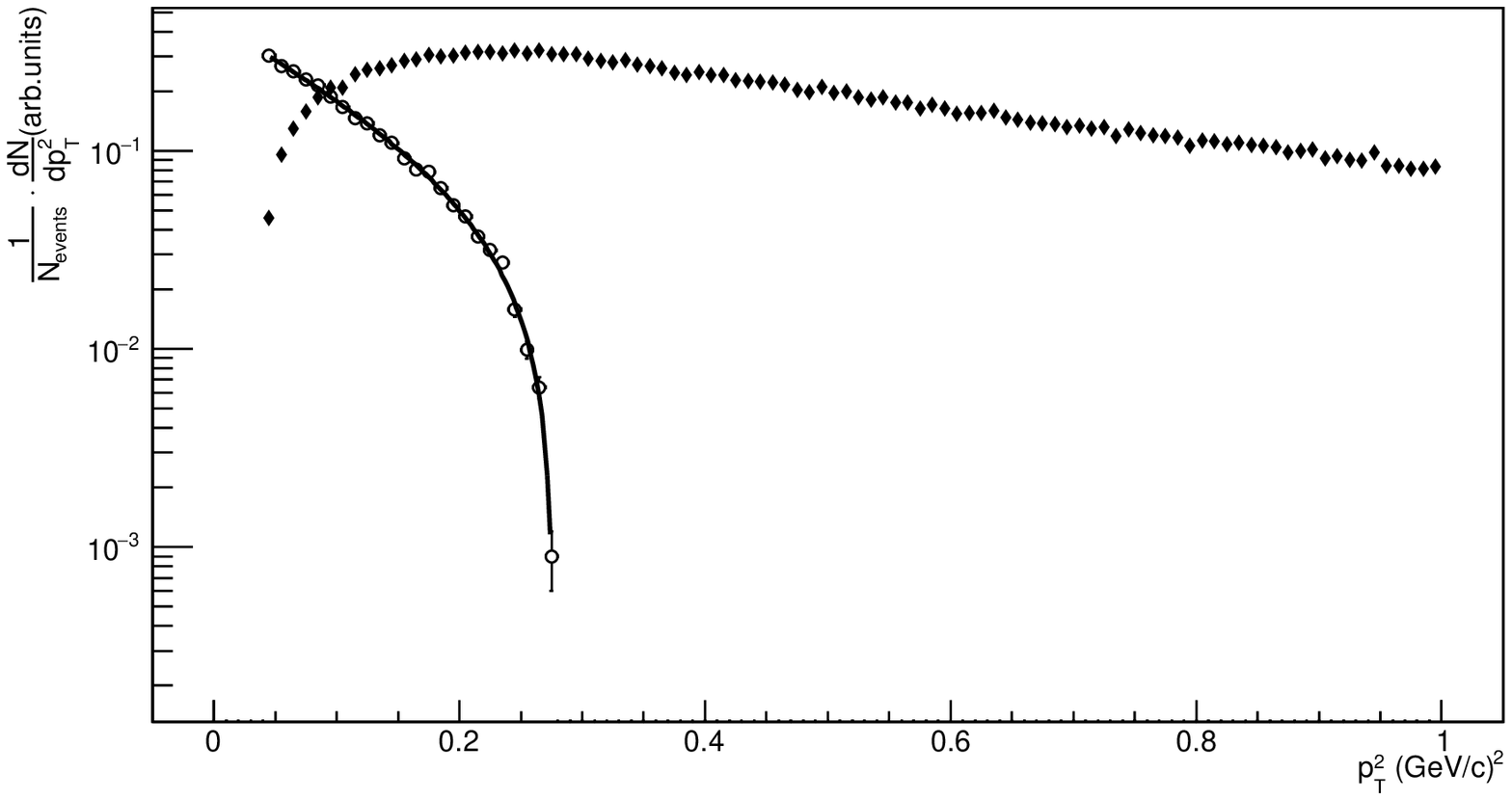}}
	\end{minipage}\\
		\begin{minipage}{.5\linewidth}
	
		\subfloat[$p_{T}^{2}$ ($p(\bar{p})$; ${\zeta}_C= 7.60$; without cuts)]{\includegraphics[scale=.43]{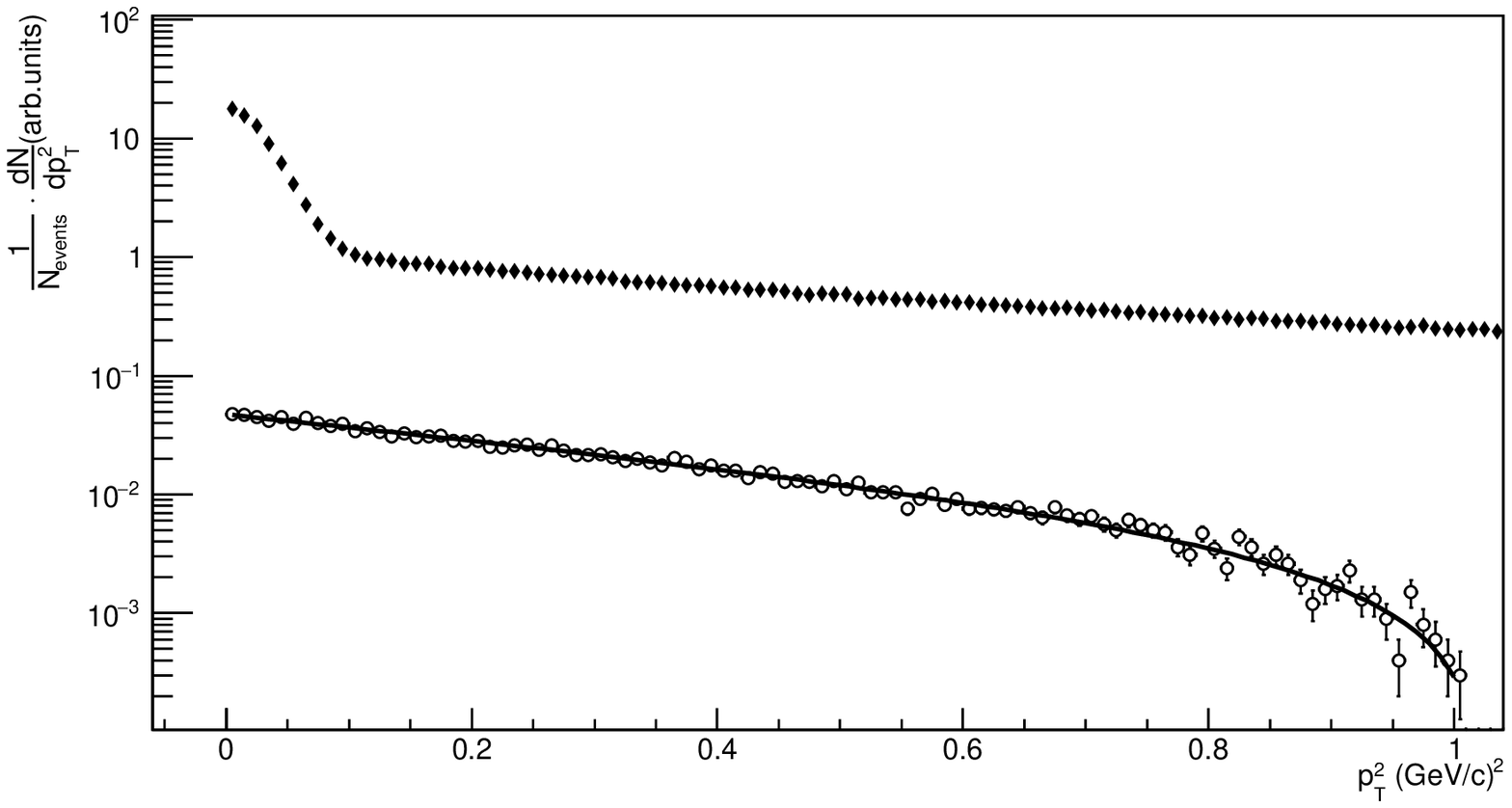}}
	\end{minipage}%
	\begin{minipage}{.5\linewidth}
	
		\subfloat[$p_{T}^{2}$ ($p(\bar{p})$; ${\zeta}_C= 7.60$; with cuts)]{\includegraphics[scale=.43]{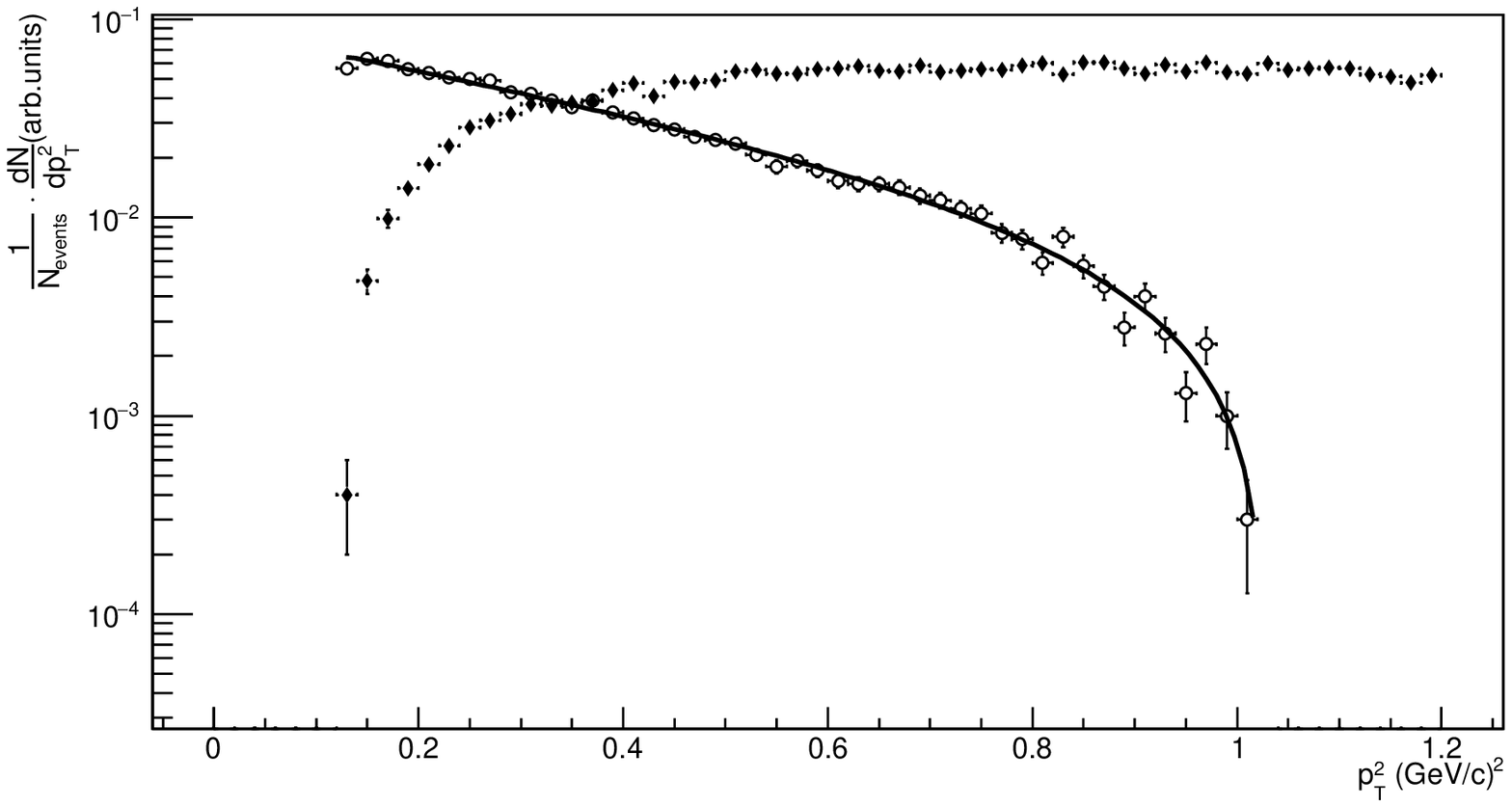}}
	\end{minipage}	
	\caption{The $p_{T}^{2}$ distribution of particles in Region-1 fitted with equation \eqref{PtSqInt} with and without kinematic and acceptance cuts are shown in open circles. Shown in filled diamonds are the corresponding distributions in Region-2.}
	\label{UrQMDPtSqPlots}
\end{figure*}
%----------------------------Main figures End----------------------------
Ten thousand Pb-Pb minimum bias collisions at $\sqrt{s} = 2.76$ TeV, simulated using the UrQMD (version 3.4) event generator \cite{UrQMD1,UrQMD2} are used for the analysis. For every such simulated collision, the inclusively produced $\pi^{\pm}$, $K^{\pm}$ and $p(\bar{p})$ are selected and a distribution of $|\zeta^{\pm}|$ is made for each of them. Since the resulting $\zeta^{\pm}$ and $cos(\theta)$  distributions of particles will be symmetric with respect to the two hemispheres, we take the absolute values of these variables to fill the distributions that are to be fitted with relevant equations. In the next step, the $|\zeta^{\pm}|$ distributions are fitted with equation \eqref{ZetaInt} and the temperature $T_{\zeta}$ is extracted. The lowest value of $\zeta$ to which it can be done successfully is taken as $|\tilde{\zeta^{\pm}}|$. We use the $\chi^2$ minimisation method as implemented in the ROOT (version 6) software package for the fitting \cite{BRUN199781}. A successful fit means that the following relation is true for it,
\begin{equation}\label{ChiSq} 
\frac{\chi^2}{n.d.f} \sim 1.0
\end{equation}
where $n.d.f$ stands for the number of degrees of freedom. The particles are then divided into two groups based on the value of the $|\tilde{\zeta^{\pm}}|$. Those particles with $|\zeta^{\pm}| > |\tilde{\zeta^{\pm}}|$ forms Region-1 and those with $|\zeta^{\pm}| < |\tilde{\zeta^{\pm}}|$ forms Region-2 in the phase space. The two regions are mutually exclusive. In the next step, the $|cos(\theta)|$ distribution of particles in Region-1 is made and is fitted with equation \eqref{CosInt}. The limit of the integral in equation \eqref{CosInt} denoted by $p_{max}$, is now governed by the $|\tilde{\zeta^{\pm}}|$ we obtained from fitting the $|\zeta^{\pm}|$ distribution. Similarly the $p_{T}^{2}$ distribution of particles in Region-1 is made and fitted with equation \eqref{PtSqInt} using the same value $|\tilde{\zeta^{\pm}}|$ for calculating $p_{z,max}$. The temperatures obtained from the fit are denoted as $T_{cos(\theta)}$ and $T_{p_{T}^2}$ respectively. In all these three cases of the fitting, the form of $f(E)$ is taken as in equation \eqref{eqboltz}. If the fits are successful, then $|\tilde{\zeta^{\pm}}|$ is taken as the final constant value $\zeta_C$ of the light front variable for the specific specie of particle. If the fits do not follow equation \eqref{ChiSq}, we repeat the above procedure with a larger value of $|\tilde{\zeta^{\pm}}|$ until the three fits are successful or the value of $|\zeta^{\pm}|$ can no longer be increased kinematically. Upon finding a $\zeta_C$ from the procedure, we say that the light front analysis scheme could select a group of thermalised particles. If the fitting cannot be done successfully for any values of the $|\tilde{\zeta^{\pm}}|$, then we say that the light front analysis scheme fails to find such a group of thermalised particles. For a full phase space analysis of the inclusive hadrons, this scheme was implemented for the simulated UrQMD collisions and we could always find a $\zeta_C$ which obeys our criterion for the three species of particles we considered. 
\subsection*{Treatment of kinematical and acceptance restrictions}
In high energy collider experiments like the ALICE placed on the Large Hadron Collider at CERN \cite{Collaboration_2008}, we only have proper access to some specific region of the phase space due to the geometry and performance of the detectors. Hence the testing of the presence of a thermalised medium in the heavy-ion collisions in such experiments with the light front analysis needs to incorporate the kinematical and acceptance restrictions in the calculations. The cut on the transverse momentum $p_{T}$ and pseudorapidity $\eta$ are the most common restrictions imposed. In this section, a method of incorporating these cuts in the light front analysis is described. The numerical values used for the cuts in our analysis as shown in Table.\ref{UrQMDrange} are inspired from the values used for the study of various species at the ALICE experiment \cite{PhysRevLett.109.252301}. 
\begin{table*}[ht!]
\begin{center}
\begin{tabular}{llll}
\noalign{\smallskip}\hline\noalign{\smallskip}
Particle & $p_{T,lowercut}$ (GeV) & $p_{T,uppercut}$ (GeV) &  $\eta $ Range\\ 
\noalign{\smallskip}\hline\noalign{\smallskip}
$ \pi^{\pm} $ & 0.10 & 3.0 & $|\eta | < 0.80 $ \\ 
$ K^{\pm}   $ & 0.20 & 3.0 & $|\eta | < 0.80 $ \\ 
$ p(\bar{p})$ & 0.35 & 4.5 & $|\eta | < 0.80 $ \\ 
\noalign{\smallskip}\hline\noalign{\smallskip}
\end{tabular}
\end{center}
\caption{$p_{T}$ and $\eta $ ranges for the light front analysis for various species.}
\label{UrQMDrange}
\end{table*}

\begin{figure*}[hbt!]\centering
\includegraphics[width=0.8\textwidth]{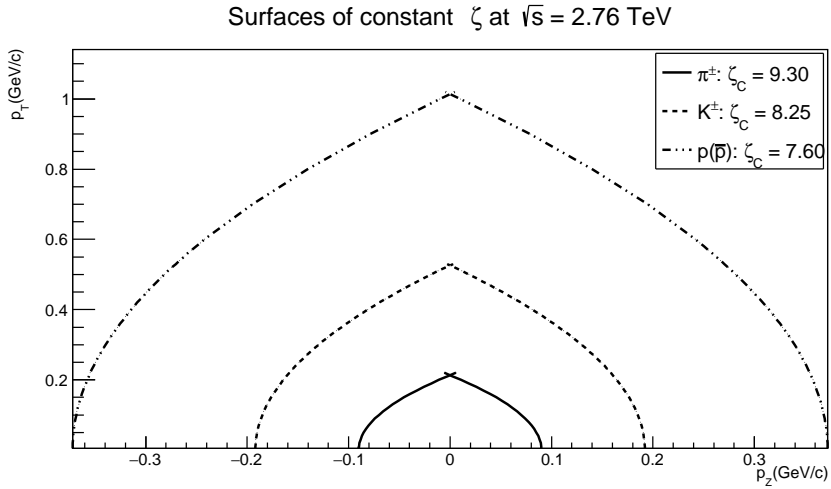}% Here is how to import EPS art
\caption{Surfaces of constant $\zeta^{\pm}$ in the Peyrou plot for UrQMD hadrons at $\sqrt{s}$ = 2.76 TeV}
\label{ConstantSurfaces}
\end{figure*}
\begin{table*}[hbt!]
\begin{center}
\begin{tabular} {llllllll}
\noalign{\smallskip}\hline\noalign{\smallskip}
\multicolumn{8}{c}{Full phase space analysis}\\
\noalign{\smallskip}\hline\noalign{\smallskip}
Species       & $\zeta_C$ & $T_{\zeta}$(MeV)& $\chi^2/ndf$&  $T_{p_T^2}$(MeV) & $\chi^2/ndf$  & $T_{cos(\theta)}$ (MeV)& $\chi^2/ndf$\\ 
\noalign{\smallskip}\hline\noalign{\smallskip}
\multicolumn{8}{c}{Full phase space analysis}\\
\noalign{\smallskip}\hline\noalign{\smallskip}

$\pi^{\pm}$   & 9.30 & 88 $\pm$ 4  & 4/22    &  87 $\pm$ 2 & 12/43  & 75 $\pm$ 1 & 103/98\\

$K^{\pm}$     & 8.25 & 213 $\pm$10 & 10/36  & 179 $\pm$ 8 & 5/26  & 166 $\pm$ 5 & 106/98\\

$p(\bar{p})$  & 7.60 & 377 $\pm$ 23 & 22/36 & 310 $\pm$ 9 & 83/99 & 288 $\pm$20 & 37/65 \\

\noalign{\smallskip}\hline\noalign{\smallskip}
\multicolumn{8}{c}{Constrained phase space analysis}\\
\noalign{\smallskip}\hline\noalign{\smallskip}
Species       & $\zeta_C$ & $T_{\zeta}$(MeV)& $\chi^2/ndf$&  $T_{p_T^2}$(MeV) & $\chi^2/ndf$  & $T_{cos(\theta)}$ (MeV)& $\chi^2/ndf$\\ 
\noalign{\smallskip}\hline\noalign{\smallskip}
$
\pi^{\pm}$   & 9.30 & 86 $\pm$ 11  & 1/13  &  93 $\pm$ 4 & 6/32  & 76 $\pm$  3 & 37/53 \\

$K^{\pm}$    & 8.25 & 206 $\pm$ 16 & 6/28  & 178 $\pm$ 12 & 4/22 & 164 $\pm$ 9 & 54/61 \\

$p(\bar{p})$ & 7.60 & 394 $\pm$ 36 & 17/30 & 341 $\pm$ 15 & 38/43  & 293 $\pm$ 13 & 89/98\\
\noalign{\smallskip}\hline\noalign{\smallskip}
\end{tabular}
\end{center}
\caption{Temperatures obtained from the light front analysis over UrQMD hadrons}
\label{UrQMDTempratureTable}
\end{table*}
Consequently, the limits of the integration in equation \eqref{ZetaInt}, equation \eqref{CosInt} and equation \eqref{PtSqInt} have to be modified. The lower limit of the integration, $p_{T,min}^{2}$ in equation \eqref{ZetaInt} is obtained by solving the following quadratic equation
\begin{equation}
p_{T}^{2} + \frac{2\sqrt{s}\xi}{tan(\theta)}p_{T} + (m^2-(\sqrt{s}\xi)^2) = 0
\end{equation}
where $\theta = 2\tan^{-1}(e^{-\eta})$ with $\eta$ = 0.8 here. After obtaining the numerical value of $p_{T,min}^{2}$ for a specific $\xi$ (subsequently  the $p_{T,max}^{2}$ from equation\eqref{ptmax}), we can compare it with the lower $p_{T,lowercut}^{2}$ (and upper $p_{T,uppercut}^{2}$) cut values of the particle under consideration respectively. We assign the larger value as the final $p_{T,min}^{2}$ (and smaller value in the case of $p_{T,max}^{2}$) for the calculation of the integral. Hence both the $p_{T}$ and $\eta$ restrictions imposed by the detector can be incorporated to perform the fitting of the $|\zeta^{\pm}|$ distribution with equation \eqref{ZetaInt}. The boundary of integration in equation \eqref{CosInt} without any cuts is given by equation \eqref{pmax} or equivalently by the solution to the following quadratic equation
\begin{equation}
\frac{sin^2(\theta)}{-2\xi\sqrt{s}}p^{2} - cos(\theta) p - \frac{m^2-(\sqrt{s}\xi)^2}{-2\xi\sqrt{s}} = 0
\end{equation}
For a specific value of $\eta$, the maximum and minimum values of the  total momentum ($p_{min}$ and $p_{max}$) possible after incorporating the cuts on the transverse momentum ($p_{T,lowercut}$ and $p_{T, uppercut}$) can be found by calculating the corresponding maximum and minimum values of the longitudinal momentum($p_{z,min}$ and $p_{z, max}$). The numerical comparison of the values obtained from the two methods, which would impart the stronger constraint, will decide the final value to be used for fitting the $cos(\theta)$ distribution of particles in Region-1 with equation \eqref{CosInt}. The cut on $\eta$ is naturally coming to the consideration via the range of $cos(\theta)$ distribution we need to fit. Finally, the cut on $p_{T}$ will naturally come to the $p_{T}^2$ distribution through the range. The upper boundary of the integration will be decided by comparing the numerical value obtained using equation \eqref{pzmax} and the one corresponding to the cut on $\eta$ for a specific value of $p_{T}^{2}$. The kinematic and acceptance restrictions are thus implemented in to the calculations of the integrals and the light front analysis was performed using this scheme for $\pi^{\pm}$, $K^{\pm}$ and $p(\bar{p})$ in the simulated UrQMD events at $\sqrt{s} = 2.76$ TeV. The results of the fits with and without the kinematical and acceptance cuts are presented in figure 1, figure 2 and figure 3 for the $\pi^{\pm}$, $K^{\pm}$ and $p(\bar{p})$ respectively. Note that the uncertainties in the distributions are statistical.The summary of the temperatures obtained from this analysis is presented in Table \ref{UrQMDTempratureTable}. The paraboloids corresponding to the constant light front variables obtained for $\pi^{\pm}$, $K^{\pm}$ and $p(\bar{p})$ are shown in figure \ref{ConstantSurfaces}.
%============================== Conclusions  ================================
\section{Conclusions and discussions} 
The light front variables introduced to study the dynamics of the production process in high energy collisions and a scheme of analysis based on these variables to investigate the existence of thermalisation in such collisions are revisited. The analysis is performed over the inclusively produced $\pi^{\pm}$, $K^{\pm}$ and $p(\bar{p})$ in Pb-Pb collisions simulated using the UrQMD event generator at $\sqrt{s} = 2.76$ TeV. The existence of a surface,  defined by a constant value of the light front variable which can select a thermalised group of particles in the phase space was confirmed through the analysis we performed. The values of temperatures we extracted from the analysis with the restricted phase space are consistent within the errors, to the corresponding temperatures from the full phase space analysis for all the three species of particles we considered. It tells us that a similar analysis using the ALICE experimental data with the kinematical and acceptance restrictions enforced by the capabilities of the detector is very much feasible to perform. One may use this method to test the formation of a thermalised medium in the ultrarelativistic heavy-ion collisions at LHC. Like in the older analysis at low energies, the temperatures extracted from the three distributions for a specific species does not converge for the values of $\zeta_C$ we found with our criterion. In the experimental analysis at ALICE, it might be interesting to know whether the three values converge for larger values of the $\zeta_C$ for a specific specie of particle, by introducing it as a criterion for the successful fitting of the three distributions. The temperatures we extracted with the UrQMD events is of significant interest. From the lattice QCD calculations, the phase transition from the hadronic degrees of freedom to the quark-gluon plasma phase happens around a temperature range of about $145 - 165$ MeV (see section 3.3 in \cite{Ratti_2018}). The temperatures we extracted for $K^{\pm}$ and $p(\bar{p})$ exceeds the Hagedorn temperature \cite{Gazdzicki2016} and the expected values for a QCD phase transition from the lattice QCD. However, the temperatures extracted for the $\pi^{\pm}$ are of the same order as in the earlier studies. A detailed investigation is required to conclude whether getting such values of temperatures with the experimental data using the light front analysis can confirm the presence of a deconfined medium in the collisions. In particular, the recent prediction about the temperature with which the charm quarks becoming a relevant degree of freedom around 250 MeV makes it interesting to extract the temperatures for the D mesons in ALICE using the light front analysis \cite{Ratti_2018, PhysRevC.98.034909}. The hydrodynamical simulations of the heavy-ion collisions at LHC energies \cite{doi:10.1142/S0218301315300106, doi:10.1146/annurev-nucl-102212-170540} suggests that the sufficiently low $p_T$ particles exhibit the characteristics of a system that has reached thermal equilibrium to which our results are in qualitative agreement.\\
\ack
I thank Prof. Marcus Bleicher (UrQMD collaboration) for the consent to use the UrQMD source code and Dr. Bharati Naik (IIT, Bombay) for helping me with the computational resources.
\appendix
\section{}\label{appendixA}
\begin{figure*}[hbt!]\centering
\includegraphics[width=0.8\textwidth]{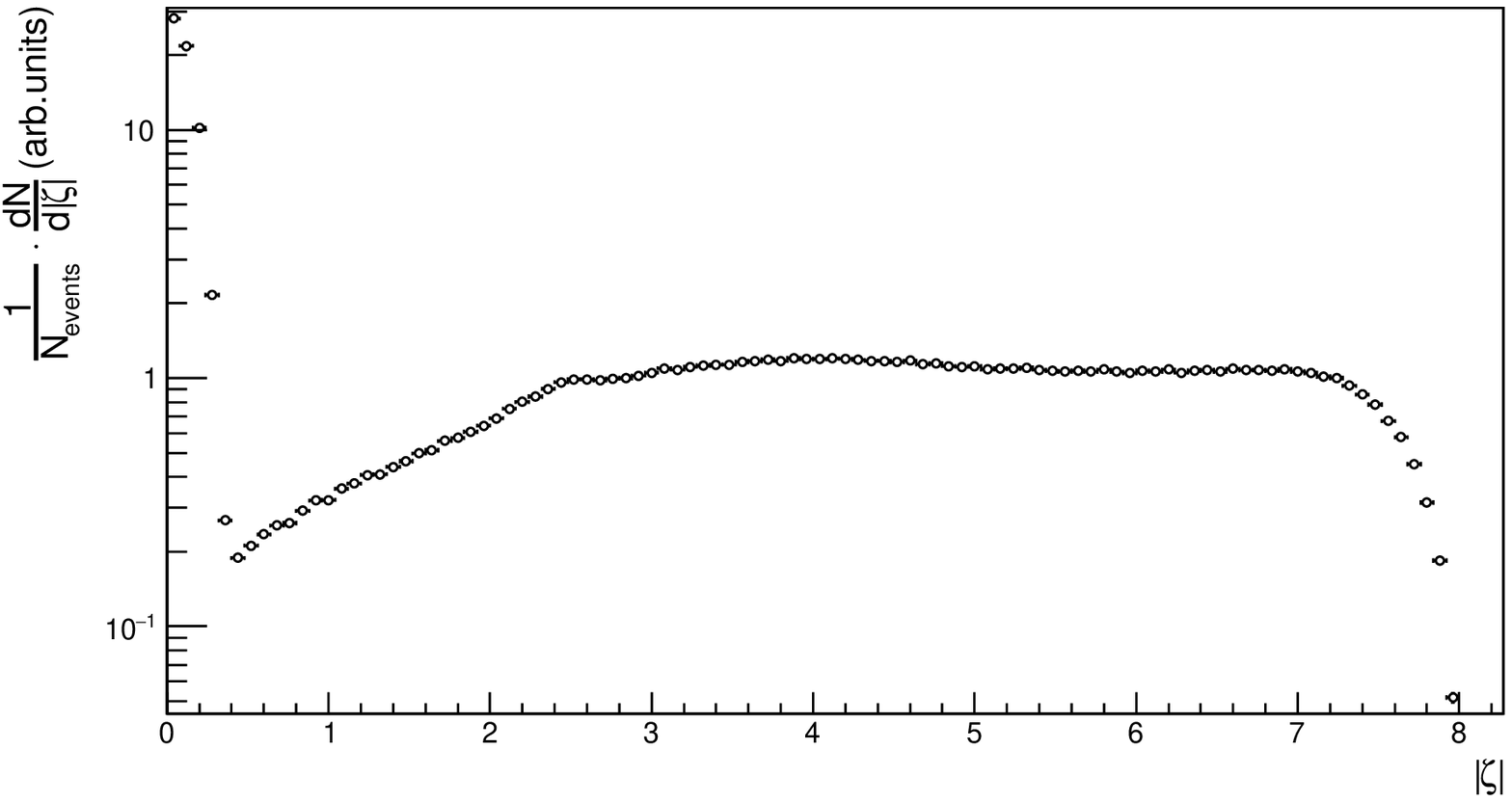}% Here is how to import EPS art
\caption{Full range $|\zeta|$ distribution of UrQMD $p(\bar{p})$ at $\sqrt{s} = 2.76$TeV}
\label{FullrangeProtonZeta}
\end{figure*}
%=========================== References ================================================
\section*{References}
\bibliographystyle{iopart-num}
\bibliography{article} 
\end{document}